# Consistent system of oscillator strengths of $A^2\Delta - X^2\Pi$ (0, 0) and $B^2\Sigma^- - X^2\Pi$ (0, 0) bands of CH molecule

Weselak[1], T., Galazutdinov[2,3], G.A., Gnaciński[4], P., Krełowski[5], J.

[1] Institute of Physics, Kazimierz Wielki University, Weyssenhoffa 11, 85-072 Bydgoszcz, Poland
e-mail:towes@gazeta.pl

[2] Astronomia, Universidad Catolica del Norte, Av. Angamos 0610, Antofagasta, Chile

[3] Pulkovo Observatory, Pulkovskoe Shosse 65, Saint-Petersburg 196140, Russia
e-mail:runizag@gmail.com

[4] University of Gdansk, ul. Wita Stwosza 57, 80-952 Gdansk, Poland
e-mail: pg@iftia9.univ.gda.pl

[5] Center for Astronomy, Nicolaus Copernicus University, Gagarina 11, Pl-87-100 Toruń, Poland
e-mail: jacek@astri.uni.torun.pl

ABSTRACT

Detailed analysis of intensity ratios of unsaturated methylidyne (CH) A–X and B–X bands suggests consistency of the recently published oscillator strengths of $A^2\Delta - X^2\Pi$ (0, 0) – 4300 Å and $B^2\Sigma^- - X^2\Pi$ (0, 0) – (3878, 3886 and 3890 Å) bands (ie. 506, 107, 320 and $213 \times 10^{-5}$ respectively). This result is based on extremely high S/N ratio spectra of 45 stars, acquired with three high-resolution spectrographs, situated in Chile: ESO LaSilla (HARPS), ESO Paranal (UVES) and Las Campanas (MIKE) and MAESTRO instrument situated in Northern Caucasus (Russia). The calculated methylidyne column densities, obtained using the consistent system of the oscillator strengths toward the observed targets, are given as well. We verify oscillator strengths of the CH+ A-X (0, 0) and (0, 1) bands at 4232 Å and 3957 Å equal to 545 and $342 \times 10^{-5}$ respectively. We also confirm the lack of correlation between abundances of neutral CH and $CH^+$ which the fact demonstrates that they are originated in different environments.

**Key words:** *ISM: molecules*

## 1. Introduction

Interstellar lines of CH molecule have been first mentioned as interstellar features of unknown origin by Beals & Blanchet (1938) and identified by McKellar (1940a, b) in spectra of OB stars due to its A–X feature centered near 4300 Å as the strongest observed interstellar line in the violet region easily accessible to photographic observations. Also the B–X system near 3886 Å is quite frequently observed though being much weaker than the A-X band. Abundances of CH molecule were proved to be very tightly correlated with those of $H_2$ molecule (Mattila 1986, Weselak et al. 2004, Liszt 2007) and also with those of OH (Weselak et al. 2009b, 2010b). The abundances of CH molecule are also well correlated with those of NH (Weselak et al. 2009c).

Laboratory analysis of the CH A–X and B–X systems in emission was accomplished in the works of Mulliken (1927) and Gerö (1941) and played important role in understanding of doublet spectra of diatomic molecules (see (Herzberg 1950) for a review). Recent analyses of lifetimes and transition probabilities of CH systems include theoretical works of Larsson & Siegbahn (1983), van Dishoeck (1986), and works based on laboratory measurements of Jeffries, Copeland & Crosley (1987), Luque & Crosley (1996a, 1996b), and Kumar et al. (1998).

From the astrophysical point of view a majority of the published column densities of CH molecule base on measurements of the equivalent widths of the strongest A–X feature; sometimes also B–X is being applied (Weselak et al. 2008b). However, in many cases the A–X feature of CH molecule is saturated which leads to differences in the column densities estimated for the same object by different authors. The column densities may also depend on the resolution and S/N ratio of the spectra used: in the case of HD 23180 the published value of column density of Federman (1982) and Crane, Lambert & Sheffer (1995) differs by a factor of 2. Apparently what matters is the problem with the setting of the baseline for the $W_\lambda$ measurements. Column densities of CH molecule base on experimentally determined values of oscillator strengths which also do not coincide in many cases (see Table 3 in Lien (1984)), and it is not an easy task to decide which of the published oscillator strengths is the most accurate and should be applied. The f-values appear in the literature in one of the two forms: the band oscillator strength ($f_{v'v''}$) and oscillator strength for individual line or blend of lines ($f_{J'J''}$). The former can be derived on the basis of radiative lifetime measurements (see formulae in (Larsson 1983)) or the ab initio calculations. The latter can be obtained from $f_{v'v''}$ through the use of the Hönl-London factor for the transition of interest. However, the conversion of $f_{J'J''}$ shows uncertainties up to the factor of 2 as suggested by Lien (1984). Our method of agreeing f-values is the same as published in previous publications (Weselak et al. 2009a, Weselak 2011). This method is very precise when based on high-quality spectra; adopted molecular parameters (f-values) are necessary to obtain precise column densities (see chapter 4 in (Gredel et al. 2011)).

This accurate assessment of the parameters of the molecular transitions under consideration requires considerable attention. Very high quality astronomical observations allow to check at least whether the existing system of the oscillator strengths is internally consistent, i.e. whether the column densities, derived from different bands of the same molecule, are identical inside the measurement error.

The formation and existence of $CH^+$ molecule, discovered in the interstellar medium (ISM) by Beals & Blanchet (1938) and identified by Douglas & Herzberg (1941), remains an unsolved problem (van Dishoeck & Black 1989, Gredel et al.

Table 1: Optical and ultraviolet transitions of CH and $CH^+$ with the published f–values for every line considered in this work (see text). The f-values and wavelengths of the CH molecule are taken from Gredel et al. (1993); these of the $CH^+$ – from Weselak et al. (2009a).

| Molecule/Ion | Transition | Band | Line | Wavelength (Å) | f-value $(10^{-5})$ |
|---|---|---|---|---|---|
| CH | $A^2\Delta - X^2\Pi$ | (0, 0) | $R_{2e}(1)+R_{2f}(1)$ | 4300.3132 | 506 |
|  | $B^2\Sigma^- - X^2\Pi$ | (0, 0) | $^PQ_{12}(1)$ | 3890.217 | 213 |
|  |  |  | $Q_2(1)+^QR_{12}(1)$ | 3886.409 | 320 |
|  |  |  | $R_2(1)$ | 3878.774 | 107 |
| $CH^+$ | $A^1\Pi - X^1\Sigma^+$ | (0, 0) | R(0) | 4232.548 | 545 |
|  |  | (1, 0) | R(0) | 3957.689 | 342 |

1993, Sheffer et al. 2008 and references therein). Based on observational spectroscopy one can argue that environments dominated by the CH molecule (i.e. regions where it is much more abundant than $CH^+$), and those where $CH^+$ dominates, are well separated in space. This conclusion is grounded in the publication of Allen (1994) which demonstrates that radial velocities of CH and $CH^+$ may differ by as much as 7.3 km/s (see also (Weselak et al. 2008a)). As a fact column densities of CH and $CH^+$ do not correlate (Sheffer et al. 2008).

In this paper we examine the ratios of equivalent widths ($W_\lambda$) of the A–X and B–X systems of CH molecule measured in the sample of 45 high-resolution spectra of OB stars of uniquely high S/N ratio, seeking unsaturated features for the analysis. We also verify oscillator strengths of the CH+ A-X (0, 0) and (0, 1) bands at 4232 Å and 3957 Å published by Weselak et al. (2009a). In Table 1 we present transitions, i.e. the identification of each feature of CH molecule to be considered, its wavelength and the individual f–value (from the recent publications); the same data are given for the two strongest bands (0,0) and (1,0) of the $CH^+$.

## 2. Observational material

We have collected the observational data, obtained using three echelle spectrographs located in Chile: UVES (ESO – Paranal), HARPS (ESO–LaSilla) and MIKE (Las Campanas) and MAESTRO instrument situated in Northern Caucasus (Mount Terskol – Russia). Only exceptionally high S/N ratio spectra are selected for our measurements.

Most of this observational material, listed in Table 2, was collected using the UVES spectrograph at ESO Paranal in Chile with the resolution equal to 80,000 in the blue range (3020 – 4980 Å) and R = 110,000 in the red range (4800 – 10,400 Å). The molecular lines of interest occupy the blue range. The data set contains spectra acquired during our observing run in March 2009 (program 082.C-0566(A)). We have also applied high signal-to-noise spectra from the ESO Archive (programs 71.C-0367(A), 076.C-0431(B)). The spectra, being averages of 50 – 150 exposures, are of exceptionally high S/N ratio which allows to measure all CH lines with the highest possible precision. The latter spectra cover only the range between 3750 and 4980 Å and thus allow measurements of A–X (0,0) and B–X (0,0) bands of CH as well as A–X (0,0) and (1,0) bands of $CH^+$.

We have added to the sample the spectra of 4 objects (HD's:147165, 147889, 147933, 179406) recorded using the HARPS spectrometer, fed with the 3.6m ESO telescope in Chile (see http://www.ls.eso.org/lasilla/sciops/3p6/harps/).
These spectra do cover the range $\sim$3800 – $\sim$6900 Å with the resolution equal to 115,000 (program: 078.C-0403(A)). As the instrument was designed to search for exoplanets, it guarantees precise wavelength measurements. The S/N ratio of the HARPS spectra is, however, lower than those from UVES instrument.

The spectra of 3 objects (HD's: 27778, 34078, 152233) were acquired using The Magellan Inamori Kyocera Echelle Instrument (Las Campanas Observatory Chile) with the highest resolution 67,000 (for more information see
http://www.Ico.cl/telescopes-information/magellan/instruments/mike). The spectra used in this work are averages of 10 exposures (HD 27778) 24 exposures (HD 34078) and 59 exposures (HD 152233).

The spectra of HD's: 23180, 24398, 24912 were obtained with MAESTRO instrument fed by the 2-m telescope of the Observatory at Peak Terskol in the Northern Caucasus (see http://www.terskol.com/telescopes/3-camera.htm). The instrument forms échelle spectra that cover the range from 3,500 to 10,100 Å with the resolution R=80,000.

To reduce all our spectra we used the standard packages: MIDAS and IRAF, and our own DECH code (Galazutdinov 1992), which provides all the standard procedures of image and spectra processing. Most of our spectra from UVES were downloaded from the archive as pipeline-reduced products (UV-Visual Echelle Spectrograph user manual) which allowed another comparison of the precision of the measured wavelengths and equivalent widths.

It is to be emphasized that the correct procedure of continuum placement is the main source of errors of the equivalent width measurements. This is especially important in the case of the B–X CH band because it is situated inside the more or less broad profile of the stellar 3889 Å Balmer line. If a weak CH features occupy an especially steep slope of the HI line profile, the baseline setting is really uncertain. In case of each spectral line the procedure presented in the publication of Weselak et al. (2009a) was performed. The error estimates were done using the formulation of Smith, Schempp & Federman (1984). The errors obtained on the basis of each fit are also presented in Table 2 and Table 3. In each case we also present calculated apparent optical depth ($\tau$) as in the publication of Weselak et al. (2011).

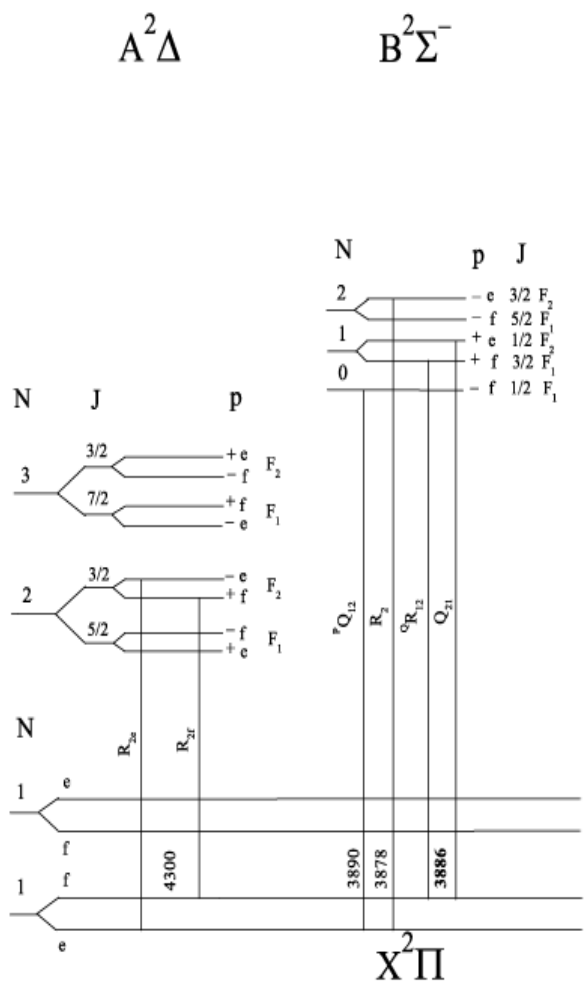

Figure 1: Term structure of the lowest rotational levels of the $A\,^2\Delta$ and $B\,^2\Sigma^-$ electronic systems of the CH molecule. Each pure rotational level N is separated into two components by spin–orbit interaction $F_1\,(J = N + 1/2)$ and $F_2\,(J = N - 1/2)$. Each J-level is split by $\Lambda$-type doubling, and is characterized by both parity (+ or –) and a letter (e or f). For more information see Herzberg (1950). With boldface we indicate 3886 feature which is observed as unresolvable blend.

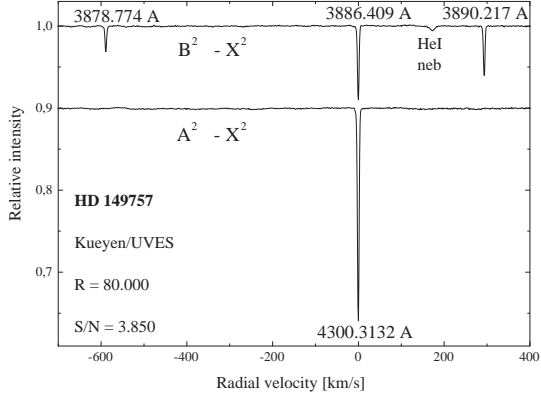

Figure 2: The CH A–X and B–X systems at 4300, 3878, 3886 and 3890 Å seen in the spectrum of HD 149757 from UVES instrument in the radial velocity scale. The stronger 4300 Å is presented at the bottom and 3878, 3886 and 3890 at the top.

Calculated apparent optical depth is equal: $\tau = 2.654 \times 10^{-15} \, f \, \lambda \, N$, where $\lambda$ is in Angstroms and column density N is in atoms cm$^{-2}$. For more information see Savage and Sembach (1991).

For this project we selected a sample of 45 reddened stars; in their spectra both A–X and/or B–X CH bands are well seen. The range covers also two strongest bands of CH$^+$: A-X (0, 0) and (1, 0) transitions near 4232 and 3957 Å respectively. Table 2 presents HD numbers, spectral type and luminosity class (Sp/L) for each star and equivalent width ($W_\lambda$s) (in mÅ) of each CH line at 3878, 3886, 3890 and 4300 Å. In several cases the weakest 3878 Å line was too weak to be reliably measured. In Table 3 we present the measured equivalent widths of two CH$^+$ bands at 4232 and 3957 Å.

The column densities of the CH molecule were derived using the profile fitting technique offered by the DECH code. Results obtained on the basis of our calculations (using the f–values from Table 1) are presented in Table 4. In the same Table we also present column densities of CH$^+$ molecule calculated using the same technique.

## 3. Results

In Fig. 1 we present the term structure of the lowest three rotational levels of the A$^2\Delta$ and B$^2\Sigma^-$ electronic systems of CH molecule. The dipole-permitted transitions from the J = 1/2 level of the X$^2\Pi$ ground state to the A$^2\Delta$ state is well seen as a blend centered at 4300.3132 Å separated by 1.43 km/s (Lien 1984). In

Table 2: The measurements of interstellar CH. Given are: star name, spectral and luminosity class, equivalent widths of the A–X 4300 Å band and B–X band (at 3886, 3890, 3878 Å) with calculated optical depth in each case (τ).

| HD | Sp/L | EW(4300) (mÅ) | τ (–) | EW(3886) (mÅ) | τ (–) | EW(3890) (mÅ) | τ (–) | EW(3878) (mÅ) | τ (–) |
|---|---|---|---|---|---|---|---|---|---|
| 23180 | B1III | 15.74 (0.18) | 0.91 (.01) | 4.23 (0.30) | 0.32 (.03) | 3.03 (0.26) | 0.23 (.02) | 1.42 (0.14) | 0.43 (.05) |
| 24398 | B1Iab | 16.17 (0.20) | 0.53 (.01) | 4.94 (0.36) | 0.38 (.04) | 3.15 (0.27) | 0.24 (.02) | 1.68 (0.21) | 0.52 (.06) |
| 24912 | O7.5Iab | 9.93 (0.28) | 0.57 (.01) | 2.88 (0.40) | 0.22 (.02) | 1.93 (0.28) | 0.14 (.01) | — | — |
| 27778 | B3V | 22.42 (0.28) | 1.53 (.02) | 7.04 (0.86) | 0.56 (.07) | 4.47 (0.78) | 0.35 (.06) | 2.83 (0.69) | 0.87 (.21) |
| 34078 | O9.5Ve | 49.12 (0.30) | 4.09 (.02) | 18.47 (1.04) | 1.51 (.08) | 11.34 (0.97) | 0.92 (.07) | 6.68 (0.79) | 2.06 (.24) |
| 35149 | B2 | 4.29 (0.07) | 0.30 (.01) | 1.12 (0.11) | 0.08 (.01) | 0.72 (0.10) | 0.05 (.01) | 0.42 (0.06) | 0.13 (.02) |
| 37903 | B1.5V | 6.40 (0.15) | 0.46 (.01) | 1.71 (0.15) | 0.13 (.01) | 1.29 (0.17) | 0.10 (.01) | 0.59 (0.04) | 0.18 (.01) |
| 52266 | O9V | 4.22 (0.35) | 0.29 (.02) | 1.51 (0.39) | 0.11 (.03) | 0.38 (0.12) | 0.03 (.01) | — | — |
| 52382 | B1Ib | 6.97 (0.13) | 0.49 (.01) | 1.84 (0.11) | 0.14 (.01) | 1.30 (0.11) | 0.10 (.01) | 0.59 (0.06) | 0.18 (.02) |
| 53974 | B0.5IV | 4.42 (0.11) | 0.31 (.01) | 1.13 (0.12) | 0.08 (.01) | 0.95 (0.15) | 0.07 (.01) | 0.58 (0.07) | 0.17 (.01) |
| 58510 | B1Ib-II | 5.16 (0.14) | 0.36 (.01) | 1.25 (0.10) | 0.09 (.01) | 0.93 (0.09) | 0.06 (.01) | 0.38 (0.07) | 0.11 (.02) |
| 73882 | O9III | 23.27 (0.11) | 1.84 (.01) | 6.87 (0.13) | 0.55 (.01) | 4.51 (0.12) | 0.36 (.01) | 2.62 (0.04) | 0.80 (.01) |
| 75149 | B3Ia | 14.72 (0.14) | 0.53 (.02) | 3.88 (0.15) | 0.19 (.01) | 1.44 (0.07) | 0.11 (.01) | 0.85 (0.08) | 0.26 (.02) |
| 76341 | B1/B2Ib | 25.58 (0.11) | 1.99 (.01) | 7.33 (0.13) | 0.58 (.01) | 4.89 (0.13) | 0.39 (.01) | 2.59 (0.07) | 0.80 (.02) |
| 91452 | B0III | 20.38 (0.13) | 1.49 (.01) | 5.96 (0.10) | 0.47 (.01) | 4.33 (0.14) | 0.32 (.01) | 1.94 (0.11) | 0.59 (.03) |
| 92964 | B2.5Iae | 10.11 (0.05) | 0.72 (.01) | 2.92 (0.08) | 0.22 (.01) | 1.84 (0.08) | 0.14 (.01) | 0.88 (0.06) | 0.27 (.02) |
| 106391 | B2Ia/Iab | 17.03 (0.67) | 0.98 (.04) | 4.87 (0.61) | 0.37 (.03) | 3.57 (0.58) | 0.27 (.02) | 1.69 (0.28) | 0.52 (.04) |
| 115363 | B2pe | 12.77 (0.05) | 1.97 (.06) | 3.44 (0.07) | 0.59 (.04) | 2.19 (0.08) | 0.37 (.02) | 1.17 (0.04) | 0.36 (.02) |
| 110432 | B1Ia | 26.57 (1.82) | 0.92 (.01) | 7.58 (1.55) | 0.25 (.01) | 4.72 (1.41) | 0.17 (.01) | — | 0.35 (.02) |
| 115842 | B0.5Ia | 22.50 (0.47) | 1.64 (.03) | 6.04 (0.32) | 0.47 (.02) | 3.45 (0.21) | 0.27 (.01) | 2.47 (0.16) | 0.76 (.05) |
| 124314 | O6V | 11.68 (0.30) | 0.67 (.03) | 2.80 (0.29) | 0.21 (.02) | 2.60 (0.33) | 0.20 (.01) | 1.18 (0.29) | 0.36 (.04) |
| 136239 | B2Iae | 21.26 (0.72) | 1.53 (.05) | 6.00 (1.48) | 0.45 (.05) | 4.59 (0.51) | 0.36 (.01) | — | — |
| 147165 | B1III | 2.96 (0.10) | 0.21 (.01) | 0.79 (0.15) | 0.06 (.01) | 0.50 (0.29) | 0.04 (.02) | — | — |
| 147889 | B2III/IV | 48.88 (0.37) | 5.23 (.04) | 20.04 (0.91) | 1.83 (.08) | 14.78 (0.95) | 1.30 (.08) | 7.43 (0.75) | 2.29 (.23) |
| 147933 | B2/B3V | 17.45 (0.26) | 1.36 (.02) | 5.18 (0.56) | 0.41 (.01) | 4.00 (0.59) | 0.32 (.01) | 1.79 (0.26) | 0.55 (.08) |
| 148184 | B2Vne | 23.54 (0.22) | 1.92 (.02) | 7.01 (0.31) | 0.56 (.02) | 4.07 (0.26) | 0.32 (.02) | 2.68 (0.33) | 0.82 (.10) |
| 148379 | B2Iap | 14.10 (0.37) | 1.01 (.02) | 3.33 (0.46) | 0.26 (.03) | — | — | 1.58 (0.04) | 0.48 (.01) |
| 148688 | B1Ia | 13.40 (0.47) | 0.97 (.03) | 3.81 (0.29) | 0.29 (.01) | 1.33 (0.23) | 0.21 (.01) | 1.19 (0.25) | 0.36 (.07) |
| 149404 | O9Ia | 22.81 (0.29) | 1.65 (.02) | 5.46 (0.29) | 0.42 (.02) | 3.95 (0.33) | 0.31 (.02) | 2.17 (0.22) | 0.67 (.06) |
| 149757 | O9V | 17.62 (0.04) | 1.36 (.01) | 5.14 (0.06) | 0.41 (.01) | 3.42 (0.06) | 0.27 (.01) | 1.82 (0.01) | 0.56 (.01) |
| 151932 | WN7h | 24.41 (0.40) | 1.80 (.03) | 6.79 (0.30) | 0.53 (.02) | 4.15 (0.35) | 0.33 (.02) | 2.49 (0.21) | 0.76 (.06) |
| 152233 | O6III | 12.73 (0.19) | 0.91 (.01) | 3.40 (0.27) | 0.26 (.03) | 1.84 (0.31) | 0.14 (.02) | 1.22 (0.34) | 0.37 (.10) |
| 152235 | B0.7Ia | 29.52 (0.07) | 2.20 (.01) | 8.03 (0.07) | 0.63 (.01) | 4.79 (0.07) | 0.37 (.01) | 2.98 (0.05) | 0.92 (.01) |
| 152236 | B0.5Ia | 20.57 (0.21) | 1.50 (.01) | 5.21 (0.18) | 0.41 (.01) | 3.57 (0.17) | 0.28 (.01) | 2.34 (0.10) | 0.72 (.03) |
| 152249 | O9Ib | 12.37 (0.44) | 0.88 (.05) | 2.58 (0.26) | 0.17 (.03) | 1.76 (0.32) | 0.01 (.01) | — | — |
| 152270 | WC+ | 13.40 (0.61) | 0.97 (.02) | 3.54 (0.48) | 0.28 (.02) | 2.64 (0.42) | 0.21 (.01) | 1.26 (0.39) | 0.39 (.12) |
| 152424 | B0Ib/II | 19.93 (0.23) | 1.44 (.01) | 4.97 (0.22) | 0.39 (.02) | 2.81 (0.20) | 0.22 (.01) | 1.63 (0.30) | 0.50 (.09) |
| 154368 | O9.5Iab | 40.86 (0.08) | 3.41 (.02) | 12.37 (0.30) | 1.02 (.02) | 8.74 (0.30) | 0.71 (.02) | 4.13 (0.11) | 1.27 (.03) |
| 154445 | B1V | 14.70 (0.24) | 1.11 (.01) | 4.14 (0.20) | 0.33 (.02) | 2.70 (0.18) | 0.21 (.01) | 1.44 (0.21) | 0.44 (.06) |
| 157038 | B4Ia | 16.32 (0.16) | 1.16 (.01) | 4.78 (0.21) | 0.37 (.01) | 2.48 (0.29) | 0.19 (.02) | 1.05 (0.20) | 0.32 (.06) |
| 163800 | O+ | 21.18 (0.25) | 1.66 (.01) | 6.66 (0.43) | 0.53 (.03) | 4.73 (0.42) | 0.39 (.02) | 2.23 (0.17) | 0.68 (.05) |
| 169454 | O+ | 31.12 (0.10) | 2.33 (.01) | 8.53 (0.10) | 0.67 (.01) | 5.60 (0.07) | 0.44 (.01) | 3.05 (0.08) | 0.94 (.02) |
| 170740 | B2V | 16.01 (0.05) | 1.16 (.02) | 4.29 (0.06) | 0.33 (.01) | 2.67 (0.08) | 0.21 (.01) | 1.44 (0.04) | 0.44 (.01) |
| 179406 | B3V | 15.62 (0.24) | 1.18 (.02) | 4.37 (0.60) | 0.34 (.05) | 3.01 (0.59) | 0.24 (.04) | 1.42 (0.14) | 0.43 (.04) |
| 210121 | B3V | 21.52 (0.63) | 1.69 (.05) | 6.94 (0.36) | 0.55 (.07) | 3.60 (0.36) | 0.29 (.06) | 2.37 (0.36) | 0.73 (.11) |

Table 3: The measurements of interstellar CH$^+$ lines. Given are: star name, spectral and luminosity class, equivalent widths of the CH$^+$ features at 4232 and 3957 Å with the calculated optical depths in each case ($\tau$).

| HD | Sp//L | EW(4232) (mÅ) | $\tau$ (–) | EW(3957) (mÅ) | $\tau$ (–) |
|---|---|---|---|---|---|
| 23180 | B1III | 6.32 (0.22) | 0.39 (0.02) | 4.04 (0.31) | 0.25 (0.02) |
| 24398 | B1Iab | 2.95 (0.14) | 0.18 (0.01) | 1.82 (0.18) | 0.11 (0.01) |
| 24912 | O7.5Iab | 21.33 (0.41) | 1.31 (0.03) | 13.23 (1.27) | 0.81 (0.10) |
| 27778 | B3V | 6.55 (0.27) | 0.46 (0.02) | 3.39 (0.28) | 0.32 (0.02) |
| 34078 | O9.5Ve | 37.67 (0.25) | 2.67 (0.16) | 22.65 (0.36) | 2.59 (0.10) |
| 35149 | B2 | 9.57 (0.07) | 0.67 (0.04) | 5.44 (0.05) | 0.52 (0.01) |
| 37903 | B1.5V | 9.68 (0.08) | 0.68 (0.04) | 5.78 (0.11) | 0.55 (0.01) |
| 52266 | O9V | 11.12 (0.18) | 0.78 (0.04) | 6.62 (0.21) | 0.63 (0.02) |
| 52382 | B1Ib | 18.57 (0.08) | 1.31 (0.08) | 10.94 (0.07) | 1.04 (0.01) |
| 53974 | B0.5IV | 11.52 (0.06) | 0.81 (0.05) | 6.46 (0.05) | 0.61 (0.01) |
| 58510 | B1Ib-II | 10.6 (0.09) | 0.75 (0.04) | 6.23 (0.07) | 0.59 (0.02) |
| 73882 | O9III | 17.66 (0.09) | 1.25 (0.07) | 10.19 (0.11) | 0.97 (0.01) |
| 75149 | B3Ia | 10.31 (1.25) | 0.73 (0.04) | 5.75 (0.06) | 0.55 (0.01) |
| 76341 | B1/B2Ib | 37.54 (0.10) | 2.66 (0.16) | 22.98 (0.08) | 2.20 (0.01) |
| 91452 | B0III | 8.82 (0.06) | 0.62 (0.03) | 4.82 (0.05) | 0.46 (0.01) |
| 92964 | B2.5Iae | 8.22 (0.09) | 0.58 (0.03) | 4.50 (0.09) | 0.43 (0.01) |
| 106391 | B2Ia/Iab | 17.90 (0.64) | 1.27 (0.05) | 9.59 (0.56) | 0.59 (0.04) |
| 115363 | B2pe | 13.8 (0.07) | 0.97 (0.06) | 7.72 (0.08) | 0.73 (0.01) |
| 110432 | B1Ia | 18.05 (1.10) | 1.10 (0.04) | 8.92 (1.16) | 0.85 (0.11) |
| 115842 | B0.5Ia | 12.86 (0.33) | 0.91 (0.05) | 7.28 (0.31) | 0.69 (0.03) |
| 124314 | O6V | 18.49 (0.22) | 1.13 (0.02) | 10.21 (0.23) | 0.62 (0.02) |
| 136239 | B2Iae | 42.93 (1.42) | 3.04 (0.18) | 23.03 (1.46) | 2.20 (0.14) |
| 147165 | B1III | 5.15 (0.13) | 0.36 (0.02) | 2.80 (0.12) | 0.26 (0.01) |
| 147889 | B2III/IV | 26.54 (0.52) | 1.88 (0.11) | 16.32 (0.50) | 1.56 (0.05) |
| 147933 | B2/B3V | 13.13 (0.24) | 0.93 (0.05) | 7.20 (0.21) | 0.68 (0.02) |
| 148184 | B2Vne | 9.57 (0.17) | 0.67 (0.04) | 5.52 (0.28) | 0.52 (0.02) |
| 148379 | B2Iap | 17.57 (0.46) | 1.24 (0.07) | 9.49 (0.51) | 0.90 (0.05) |
| 148688 | B1Ia | 22.99 (0.30) | 1.62 (0.10) | 13.32 (0.34) | 1.27 (0.03) |
| 149404 | O9Ia | 37.04 (0.30) | 2.62 (0.16) | 21.12 (0.32) | 2.02 (0.03) |
| 149757 | O9V | 23.28 (0.04) | 1.65 (0.10) | 13.54 (0.03) | 1.29 (0.01) |
| 151932 | WN7h | 13.04 (0.26) | 0.92 (0.05) | 7.12 (0.27) | 0.68 (0.02) |
| 152233 | O6III | 21.38 (0.24) | 1.51 (0.09) | 12.24 (0.43) | 1.17 (0.04) |
| 152235 | B0.7Ia | 40.84 (0.08) | 2.89 (0.17) | 24.88 (0.07) | 2.38 (0.01) |
| 152236 | B0.5Ia | 18.41 (0.20) | 1.34 (0.08) | 9.47 (0.18) | 0.90 (0.02) |
| 152249 | O9Ib | 16.29 (0.22) | 1.15 (0.07) | 9.17 (0.34) | 0.87 (0.03) |
| 152270 | WC+ | 17.87 (0.45) | 1.26 (0.07) | 9.99 (0.47) | 0.95 (0.04) |
| 152424 | B0Ib/II | 34.54 (0.35) | 2.44 (0.15) | 20.21 (0.18) | 1.93 (0.02) |
| 154368 | O9.5Iab | 19.56 (0.23) | 1.38 (0.08) | 10.68 (0.24) | 1.02 (0.02) |
| 154445 | B1V | 17.39 (0.24) | 1.23 (0.07) | 10.02 (0.21) | 0.96 (0.02) |
| 157038 | B4Ia | 48.54 (0.24) | 3.43 (0.21) | 27.72 (0.27) | 2.65 (0.02) |
| 163800 | O+ | 13.5 (0.33) | 0.95 (0.06) | 7.72 (0.32) | 0.73 (0.03) |
| 169454 | O+ | 18.07 (0.12) | 1.28 (0.07) | 9.65 (0.13) | 0.92 (0.01) |
| 170740 | B2V | 13.92 (0.08) | 0.98 (0.06) | 7.99 (0.07) | 0.76 (0.01) |
| 179406 | B3V | 3.57 (0.30) | 0.25 (0.01) | 1.75 (0.31) | 0.16 (0.03) |
| 210121 | B3V | 9.47 (0.23) | 0.67 (0.04) | 5.62 (0.20) | 0.54 (0.02) |

Figure 1 we also mark four possible transitions from the $X^2\Pi$ ground state to $B^2\Sigma^-$ electronic system: the line at 3886 Å is also the blend with the separation of 1.28 km/s.

To obtain column density we used the relation presented by Herbig (1968) which gives proper column densities when the observed lines are unsaturated:

$$N = 1.13 \times 10^{20} W_\lambda / (\lambda^2 f), \qquad (1)$$

where $W_\lambda$ and $\lambda$ are in Å and column density in cm$^{-2}$.

The intensity ratio of two unsaturated spectral lines is equal to:

$$\frac{W_{\lambda_1}}{W_{\lambda_2}} = \frac{f_1}{f_2} \frac{\lambda_1^2}{\lambda_2^2}, \qquad (2)$$

where $W_{\lambda_1}, W_{\lambda_2}$ are the equivalent widths; $f_1$, $f_2$ – the oscillator strengths (f–values) and $\lambda_1$, $\lambda_2$ – the wavelengths of the two lines under consideration. Equation (2) describes the ratio of equivalent widths of two different lines of the same molecule and originated from the same level in relation to their wavelengths and f-values. For more information see Larsson (1983).

In Fig. 2 we present the 4300, 3878, 3886 and 3890 Å lines of CH molecule in the spectrum of HD 149757. The spectrum is the average of 150 exposures and its S/N ratio is close to 4,000. One can notice the weak HeI nebular line, recently described by Galazutdinov & Krełowski (2012). The Λ–doubling of the two strongest transitions in both bands is not seen at the UVES resolution (R = 80,000), because it is too low. See Crane Lambert and Sheffer (1995) for detailed study of the Λ–doubling in CH A–X and B–X bands at higher resolution.

The equivalent width ratio of the A-X 4300 Å and CH B-X 3886 Å bands, originated from the same level, should be equal to 3.87 (using the equation (2) with the applied oscillator strengths equal to 506 and 320 $\times 10^{-5}$ taken from the publications of Larsson & Siegbahn (1983) and Lien (1984), respectively. As it is seen in Table 4 column density obtained on the basis CH A-X 4300 Å is two times higher than value from CH B-X 3886 Å band. As a result equivalent width ratio of the CH 4300 and 3886 bands obtained from Eq. (2) should be multiplied by a factor of 2. This relation is represented by the broken line in Fig. 3. Figure 4 presents spectra of four stars (HD 152424, HD 152236, ρ Oph and ζ Oph) depicted in Fig. 3. One can conclude that the 4300 Å line is slightly saturated in such objects as HD 149757 and HD 147933. Thus, the CH A–X line likely starts getting saturated when $W_\lambda$ is higher than 17 mÅ in the absence of Doppler splitting. This value is close to 20 mÅ previously proposed by van Dishoeck & Black (1989). It is to be emphasized that, as a result of the Doppler splitting, saturation effects in CH 4300 Å line toward HD's 152424 and 152236 are absent despite the higher values of the equivalent widths.

The B-X transition of CH molecule was first measured by Adams in the spectrum of ζ Oph (McKellar 1940b). The same band of CH molecule have been observed with a higher precision by (Herbig 1968) also in the case of HD 149757.

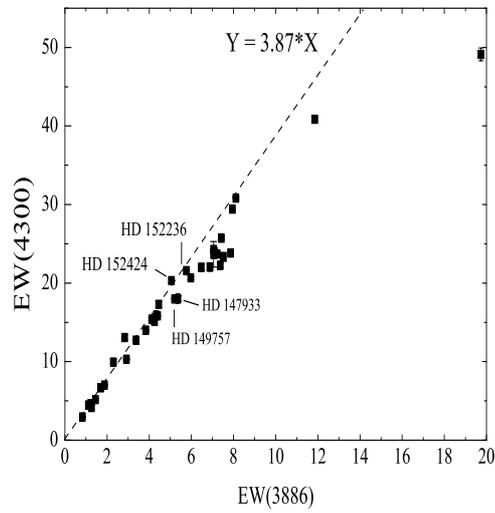

Figure 3: The correlation plot between equivalent widths of the bands at 4300 and 3886 Å. Theoretical intensity ratio is marked with the dashed line. Relation of both observed equivalent widths begins to deviate from theoretical (broken) line as a result of saturation.

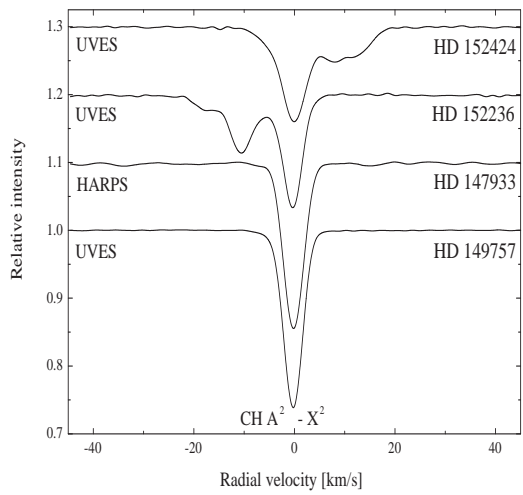

Figure 4: The spectra of four stars with equivalent widths of CH line at 4300 Å close to 20 mÅ; HD's 152424 and 152236 show no saturation because of the Doppler splitting. In case of HD's 149757 and 147933 weak saturation effects are probably seen. HD's 154368, 147889 and 34078 are heavily saturated – see the calculated apparent optical depths in Table 2.

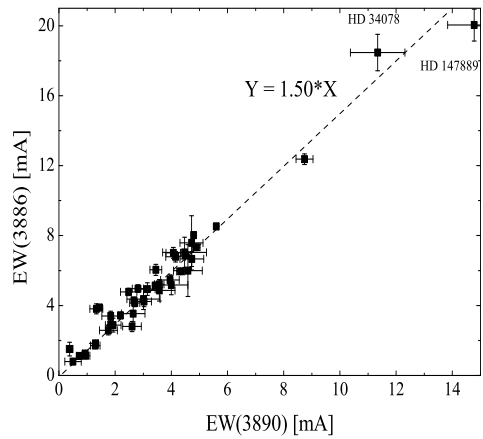

Figure 5: Correlation plot between equivalent widths of CH B-X band at 3886 Å and 3890 band. The average equivalent width ratio seems consistent with that calculated from the published oscillator strengths using equation (1) – see Table 1.

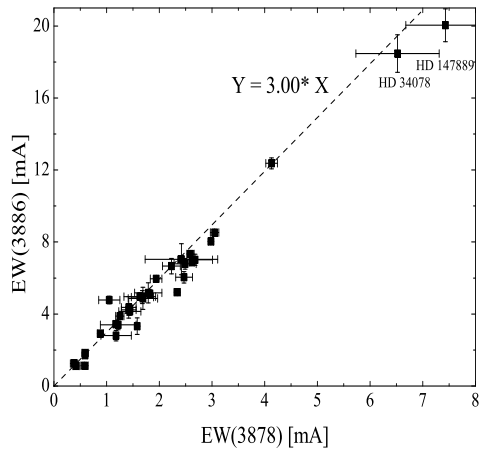

Figure 6: Correlation plot between equivalent widths of CH B-X band at 3886 Å and 3878 band. The average equivalent width ratio seems consistent with that calculated from the published oscillator strengths using equation (1) – see Table 1. Interstellar lines of CH at 3886 Å toward HD 34078 and HD 147889 are saturated – see the calculated optical depths in Table 2.

In Table 5 we compare previous measurements of the CH B–X band (not taken by the authors) with those presented in this paper. It is well seen that cited mea-

Table 4: CH and CH$^+$ column densities (in $10^{12}$ cm $^{-2}$) obtained on the basis of DECH profile fitting procedure. With $^a$ we mark column density of CH$^+$ taken from publication of Weselak et al. 2009b (in this case CH$^+$ lines at 4232 and 3957 Å were saturated).

| Star | CH A-X 4300 Å | +/- | CH B-X 3886 Å | +/- | CH B-X 3890 Å | +/- | CH B-X 3878 Å | +/- | CH$^+$ 4232 Å | +/- | CH$^+$ 3957 Å | +/- |
|---|---|---|---|---|---|---|---|---|---|---|---|---|
| HD 23180 | 19.00 | 0.22 | 9.89 | 0.70 | 10.62 | 0.91 | 19.94 | 1.97 | 7.31 | 0.25 | 8.52 | 0.65 |
| HD 24398 | 19.52 | 0.24 | 11.54 | 0.84 | 11.04 | 0.95 | 23.59 | 2.95 | 3.41 | 0.18 | 3.84 | 0.38 |
| HD 24912 | 11.99 | 0.34 | 6.73 | 0.94 | 6.76 | 0.98 | — | — | 24.69 | 0.47 | 27.91 | 2.68 |
| HD 27778 | 29.46 | 0.37 | 16.90 | 2.06 | 16.20 | 2.80 | 39.73 | 9.69 | 7.58 | 0.31 | 7.15 | 0.59 |
| HD 34078 | 70.80 | 0.44 | 45.80 | 2.58 | 41.90 | 3.58 | 93.78 | 11.09 | — | | 57.28 | 2.25$^a$ |
| HD 35149 | 5.27 | 0.07 | 2.64 | 0.26 | 2.56 | 0.37 | 5.90 | 0.84 | 11.08 | 0.08 | 11.48 | 0.11 |
| HD 37903 | 7.98 | 0.19 | 4.05 | 0.36 | 4.63 | 0.61 | 8.28 | 0.56 | 11.20 | 0.09 | 12.19 | 0.23 |
| HD 52266 | 5.16 | 0.42 | 3.54 | 0.92 | 1.36 | 0.45 | — | | 12.87 | 0.21 | 13.96 | 0.44 |
| HD 52382 | 8.62 | 0.16 | 4.33 | 0.26 | 4.65 | 0.41 | 8.28 | 0.84 | 21.49 | 0.09 | 23.08 | 0.15 |
| HD 53974 | 5.40 | 0.13 | 2.66 | 0.29 | 3.39 | 0.54 | 8.14 | 0.98 | 13.33 | 0.07 | 13.63 | 0.11 |
| HD 58510 | 6.32 | 0.25 | 2.94 | 0.20 | 3.01 | 0.32 | 5.33 | 0.98 | 12.27 | 0.10 | 13.14 | 0.15 |
| HD 73882 | 31.96 | 0.16 | 16.73 | 0.30 | 16.49 | 0.43 | 36.78 | 0.56 | 20.44 | 0.10 | 21.50 | 0.23 |
| HD 75149 | 9.14 | 0.36 | 5.75 | 0.06 | 5.10 | 0.22 | 11.93 | 1.12 | 11.93 | 1.45 | 12.13 | 0.13 |
| HD 76341 | 34.43 | 0.15 | 17.70 | 0.31 | 17.78 | 0.46 | 36.36 | 0.98 | 43.45 | 0.12 | 48.48 | 0.17 |
| HD 91452 | 25.76 | 0.17 | 14.36 | 0.24 | 14.82 | 0.39 | 27.24 | 1.54 | 10.21 | 0.07 | 10.17 | 0.11 |
| HD 92964 | 12.52 | 0.06 | 6.86 | 0.12 | 6.55 | 0.28 | 12.35 | 0.84 | 9.51 | 0.10 | 9.49 | 0.19 |
| HD 106391 | 20.56 | 0.81 | 11.38 | 1.43 | 12.51 | 0.56 | 23.73 | 3.93 | 20.72 | 0.74 | 20.23 | 1.18 |
| HD 110432 | 16.02 | 0.06 | 7.72 | 0.08 | 7.83 | 0.15 | 16.43 | 0.56 | 15.97 | 0.08 | 16.28 | 0.17 |
| HD 115363 | 34.20 | 1.10 | 18.02 | 1.16 | 16.99 | 1.23 | — | | 20.89 | 1.27 | 18.82 | 2.45 |
| HD 115842 | 28.40 | 0.58 | 14.31 | 0.52 | 12.42 | 0.39 | 34.68 | 2.25 | 14.88 | 0.38 | 15.36 | 0.65 |
| HD 124314 | 14.10 | 0.36 | 6.54 | 0.68 | 9.11 | 0.33 | 16.57 | 4.07 | 21.40 | 0.25 | 21.54 | 0.49 |
| HD 136239 | 26.48 | 0.96 | 14.16 | 1.56 | 16.41 | 3.50 | — | | 49.69 | 1.64 | 48.58 | 3.08 |
| HD 147165 | 3.63 | 0.25 | 1.84 | 0.12 | 1.77 | 1.05 | — | | 5.96 | 0.15 | 5.91 | 0.25 |
| HD 147889 | 90.64 | 0.69 | 55.47 | 2.60 | 59.14 | 3.81 | 104.31 | 10.53 | 30.72 | 0.60 | 34.43 | 1.05 |
| HD 147933 | 23.56 | 0.35 | 12.57 | 0.14 | 14.53 | 0.24 | 25.13 | 3.65 | 15.20 | 0.28 | 15.19 | 0.44 |
| HD 148184 | 33.23 | 0.32 | 17.04 | 0.76 | 14.84 | 0.96 | 37.62 | 4.63 | 11.08 | 0.20 | 11.64 | 0.59 |
| HD 148379 | 17.36 | 0.16 | 7.82 | 1.10 | — | | 22.18 | 0.56 | 20.34 | 0.53 | 20.02 | 1.08 |
| HD 148688 | 16.90 | 0.60 | 9.03 | 0.34 | 9.42 | 0.71 | 16.71 | 3.51 | 26.61 | 0.35 | 28.10 | 0.82 |
| HD 149404 | 28.66 | 0.36 | 12.91 | 0.67 | 14.17 | 1.17 | 30.46 | 3.09 | 42.87 | 0.35 | 44.55 | 0.68 |
| HD 149757 | 23.52 | 0.05 | 12.42 | 0.14 | 12.43 | 0.21 | 25.55 | 0.14 | 26.94 | 0.05 | 28.56 | 0.06 |
| HD 151932 | 31.25 | 0.61 | 16.29 | 0.72 | 15.00 | 1.24 | 34.96 | 2.95 | 15.09 | 0.30 | 15.02 | 0.57 |
| HD 152233 | 15.76 | 0.24 | 8.02 | 0.86 | 6.55 | 0.97 | 17.13 | 4.77 | 24.74 | 0.28 | 25.82 | 0.91 |
| HD 152235 | 38.19 | 0.26 | 19.18 | 0.17 | 17.07 | 0.25 | 41.84 | 0.70 | 47.27 | 0.09 | 52.48 | 0.15 |
| HD 152236 | 25.99 | 0.25 | 12.34 | 0.18 | 12.82 | 0.59 | 32.85 | 1.40 | 21.31 | 0.23 | 19.98 | 0.38 |
| HD 152249 | 15.30 | 0.79 | 5.11 | 0.95 | — | | — | | 18.85 | 0.25 | 19.34 | 0.72 |
| HD 152270 | 16.80 | 0.45 | 8.43 | 0.47 | 9.47 | 0.55 | 17.69 | 5.48 | 20.68 | 1.02 | 21.07 | 0.99 |
| HD 152424 | 24.98 | 0.26 | 11.77 | 0.56 | 10.07 | 0.56 | 22.88 | 4.21 | 39.98 | 0.41 | 42.63 | 0.38 |
| HD 154368 | 59.16 | 0.44 | 30.80 | 0.76 | 32.44 | 1.15 | 57.98 | 1.54 | 22.64 | 0.27 | 22.53 | 0.23 |
| HD 154445 | 19.22 | 0.30 | 9.90 | 0.47 | 9.74 | 0.67 | 20.22 | 2.95 | 20.13 | 0.28 | 21.14 | 0.44 |
| HD 157038 | 20.09 | 0.20 | 11.22 | 0.46 | 8.86 | 1.05 | 14.74 | 2.81 | 56.18 | 0.28 | 58.47 | 0.57 |
| HD 163800 | 28.74 | 0.33 | 16.20 | 1.04 | 17.65 | 1.22 | 31.31 | 2.39 | 15.62 | 0.38 | 16.28 | 0.68 |
| HD 169454 | 40.44 | 0.20 | 20.54 | 0.23 | 20.35 | 0.20 | 42.82 | 1.12 | 20.91 | 0.14 | 20.36 | 0.27 |
| HD 170740 | 20.14 | 0.46 | 10.16 | 0.17 | 9.57 | 0.20 | 20.22 | 0.56 | 16.11 | 0.09 | 16.85 | 0.15 |
| HD 179406 | 20.45 | 0.36 | 10.50 | 1.50 | 10.91 | 2.14 | 19.94 | 1.97 | 4.13 | 0.35 | 3.69 | 0.65 |
| HD 210121 | 29.40 | 0.85 | 16.86 | 2.30 | 13.13 | 2.73 | 33.27 | 5.05 | 10.96 | 0.27 | 11.86 | 0.42 |

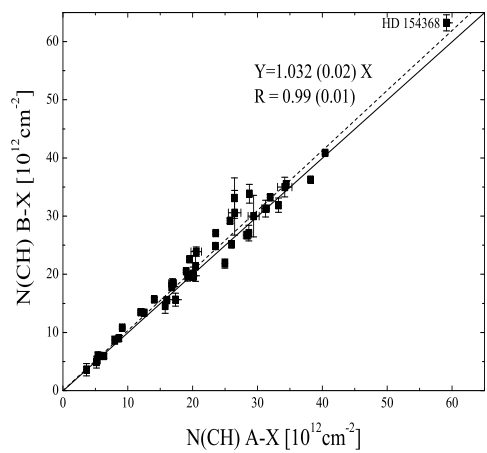

Figure 7: The relation between column density of CH molecule obtained from A –X and B – X bands. Column density derived from A – X band was obtained from unsaturated CH A–X band at 4300 Å (HD's 34078 and 147889 were excluded). Column density derived from B – X band was taken as a sum of those obtained on the basis of bands at 3886 and 3890 Å. More information see Lien (1984). Note a very tight correlation between both column densities (dashed line). With the solid line we present the perfect relation between column densities obtained on the basis of both A–X and B–X bands.

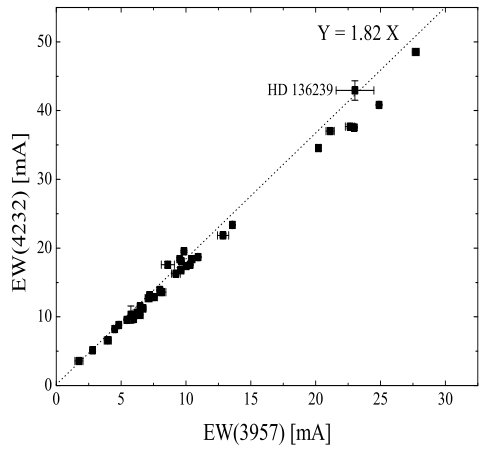

Figure 8: The relation between equivalent widths of the two strong bands of $CH^+$ at 4232 and 3957 Å. The relation resulting from equation (1) (broken line) confirms the f–values presented in Table 1. HD 136239 shows evident Doppler splitting which makes the features unsaturated.

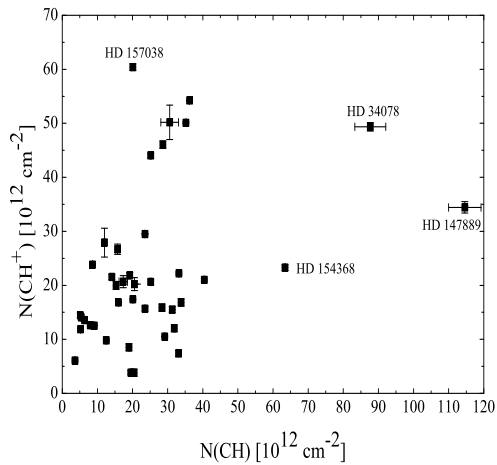

Figure 9: Lack of any relation between column densities of interstellar CH and $CH^+$. Apparently both species represent another environments.

surements disagree with ours. We trust our ones because in most cases they base on spectra of higher quality. Some measurements of the B – X CH band were also published by Federman et al. (1994) but their precision seems to be quite low.

To check whether the recently published f-values of CH B–X transitions are correct we present (in Fig. 5) the relations between equivalent widths of the 3886 Å and 3890 Å CH B–X band lines. As it is seen in Fig. 5 this ratio matches the value expected from equation (1) i.e. 1.50; this confirms both the high precision of our measurements of equivalent widths and the used f–values. This result is also consistent with the one previously published by Weselak et al. (2011).

In Fig. 6 we also present correlation plot between equivalent widths of 3886 Å and 3878 lines of the CH B–X band. Also in this case the equivalent width ratio (3.00) is close to that following the equation (1) on the basis of f–values recently published in the literature (see Table 1).

In Fig. 7 we present the relation between column densities of the CH molecule, derived as the sum of the B–X 3886 and 3890 Å lines and that obtained from unsaturated 4300 Å band using the oscillator strength equal to $506 \times 10^{-5}$ for the latter. The perfect agreement is evident. All the above figures clearly suggest that the oscillator strengths, listed in Table 1, are consistent. The system of f–values, given in Table 1, can be recommended for future investigations.

In Fig. 8 we present the relation between equivalent widths of two strong bands of $CH^+$ at 4232 and 3957 Å. The result is close to that presented in the publication of Weselak et al. (2009a), though based on spectra of much higher precision, confirming the consistency of the f–values system published in the Table 1. Thus the system of f–values of molecular transitions, listed in Table 1, is internally consistent as confirmed by the high precision observations.

Finally, in Fig. 9, we present a relation between column densities, reliably derived from our high quality spectra, using the consistent system of f–values, of interstellar CH and $CH^+$ species. The apparent lack of any correlation suggests their different reaction pathways as previously suggested by Federman (1982). It is important that the latter suggestion is now confirmed using very high S/N spectra.

## 4. Conclusions

The above considerations led us to infer the following conclusions:

1. The sample of 45 high resolution, high S/N (2,000 – 4,000) ratio spectra, allowed to confirm convincingly the recently published estimate of the oscillator strength of CH A–X transition as equal to $506 \times 10^{-5}$ which is consistent with the value of Larsson and Siegbahn (1983). This result is obtained on the assumption that f-values of the CH B-X system (at 3878, 3886 and 3890 Å), i.e. 107, 320 and $213 \times 10^{-5}$ respectively (being internally consistent) are correct. All the above values are recommended for future determinations of CH column density.

2. We confirm the consistency of A – X (0, 0) and (1, 0) transitions of $CH^+$ f-values, recently published by Weselak et al. (2009a), using the spectra of the highest quality ever used.

3. We confirm the lack of correlation between column densities of interstellar CH and $CH^+$. These two "relative" species are evidently formed in different reaction pathways and occupy environments characterized by different physical parameters.

Table 1 contains the set of identifications, precise wavelengths and f-values of CH and $CH^+$ bands carefully checked using the high class observational material and being consistent with experimental and theoretical results.

More efforts in this area is to be done in the case of known visible transitions of diatomic molecules. These data allow to adjust column densities of CH and $CH^+$ molecules toward any observed target. The history of varying estimations of f–values of the CH A–X and B–X transitions, as presented by Lien (1984), seemingly came to the convergent point.

**Acknowledgements.** JK and TW acknowledge the financial support of the Polish National Center for Science during the period 2011 - 2014 (grant UMO-2011/01/BST2/05399). GG acknowledges the support of Chilean fund FONDECYT-regular (project 1120190). We are grateful for the assistance of the ESO and Las Campanas observatories staff members.

Table 5: Published equivalent widths of the CH B–X (0, 0) transitions toward six stars presented in advance of print and compared with those presented in this paper (tha last column). References in each column : H68 – Herbig (1968), CW86 – Cardelli and Wallerstein (1986), DFL84 – Danks, Federman, Lambert (1984), GvDB93 – Gredel, van Dishoeck and Black (1993).

| HD number | H68 | CW86 | DFL84 | GvDB93 | This work |
|---|---|---|---|---|---|
| | EW(3878), EW(3886), EW(3890) mÅ(err), mÅ(err), mÅ(err) | EW(3878), EW(3886), EW(3890) mÅ(err), mÅ(err), mÅ(err) | EW(3878), EW(3886), EW(3890) mÅ(err), mÅ(err), mÅ(err) | EW(3878), EW(3886), EW(3890) mÅ(err), mÅ(err), mÅ(err) | EW(3878), EW(3886), EW(3890) mÅ(err), mÅ(err), mÅ(err) |
| 110432 | | | | ≤1.5, 3.8(0.7), 2.6(0.7) | –, 7.58(1.55), 4.72(1.41) |
| 147889 | | 5.4(1.1), 15.6(1.5), 9.4(1.4) | | | 7.43(0.75), 20.04(0.91), 14.78(0.95) |
| 148184 | | | 3.0(0.4), 6.1(0.3), 4.4(0.4) | | 2.68(0.82), 7.01(0.31), 4.07(0.26) |
| 149757 | 3:, 5.9, 5.6 | | 1.6(0.4), 5.0(0.4), 3.3(0.4) | | 1.82(0.01), 5.14(0.06), 3.42(0.06) |
| 154368 | | | | 4.0(1.0), 10.5(1.4), 6.0(1.0) | 4.13(0.11), 12.37(0.30), 8.74(0.30) |
| 169454 | | | | 3.0(0.6), 5.4(1.8), 4.5(0.7) | 3.05(0.08), 8.53(0.10), 5.60(0.07) |